\documentclass[onecolumn,showpacs,nobibnotes,nofootinbib]{revtex4}
\usepackage{amsmath,amssymb,bm,textcomp}
\usepackage{epsfig}
\usepackage{graphicx}
\usepackage{slashed}
\usepackage{psfig}

\def\lsim{\raise0.3ex\hbox{$<$\kern-0.75em\raise-1.1ex\hbox{$\sim$}}}
\def\gsim{\raise0.3ex\hbox{$>$\kern-0.75em\raise-1.1ex\hbox{$\sim$}}}

\def\pom{{I\!\!P}}
\newcommand{\be}{\begin{equation}}
\newcommand{\ee}{\end{equation}}
\def\alphaem{\alpha_{em}}

\def\beq{\begin{equation}}
\def\eeq{\end{equation}}
\def\beqa{\begin{eqnarray}}
\def\eeqa{\end{eqnarray}}

\newcommand{\rr}{\mbox{\boldmath $r$}}
\newcommand{\rrn}{\mbox{$r$}}

\newcommand{\rb}{\mbox{\boldmath $b$}}

\def\gappeq{\mathrel{\rlap {\raise.5ex\hbox{$>$}}
{\lower.5ex\hbox{$\sim$}}}}

\def\lappeq{\mathrel{\rlap{\raise.5ex\hbox{$<$}}
{\lower.5ex\hbox{$\sim$}}}}

\def\Toprel#1\over#2{\mathrel{\mathop{#2}\limits^{#1}}}

\def\pom{{I\!\!P}}

\begin{document}

\title{Diffractive deep inelastic scattering in an AdS/CFT inspired model: A phenomenological
study
} 
\author{  M.A. Betemps $^{1,2}$, V.P. Gon\c{c}alves $^2$, J. T. de Santana Amaral $^2$}
\affiliation{$^1$ Conjunto Agrot\'ecnico Visconde da Gra\c{c}a (CAVG) \\Universidade Federal de Pelotas,\\
Av. Ildefonso Sim\~oes Lopes, 2791\\ CEP 96060-290, Pelotas, RS, Brazil \\
$^2$ High and Medium Energy Group (GAME), \\
Instituto de F\'{\i}sica e Matem\'atica,  Universidade
Federal de Pelotas, 
Caixa Postal 354, CEP 96010-900, Pelotas, RS, Brazil}
\begin{abstract}
The analytical treatment of the nonperturbative QCD dynamics is one of main
open questions of the strong interactions. Currently, it is only possible to get some qualitative information about this regime considering other QCD-like theories, as for example the $N=4$ super Yang-Mills (SYM), where one can perform calculations in the nonperturbative  limit of large 't Hooft coupling using the Anti-de Sitter space/Conformal field theory (AdS/CFT). 
Recently, the high energy scattering amplitude was calculated in the AdS/CFT approach, applied to deep inelastic scattering (DIS) and  confronted with the $F_2$ HERA data. 
In this work we extend the nonperturbative AdS/CFT inspired model for diffractive processes and  compare its predictions with a perturbative approach based on the Balitsky - Kovchegov (BK) equation. We demonstrate that the AdS/CFT inspired model is not able to describe the current $F_2^{D(3)}$ HERA data and predicts a similar behavior to that from BK equation in the range $10^{-7} \lesssim x_{I\!\!P} \lesssim 10^{-4}$. At smaller values of $x_{I\!\!P}$ the diffractive structure function is predicted to be energy independent.
\end{abstract}
\pacs{12.38.Aw, 13.60.Hb, 11.25.Sq}
\keywords{Diffractive deep inelastic scattering, AdS/CFT correspondence, QCD dynamics}
\maketitle
\vspace{1cm}

\section{Introduction}
The physics of high density QCD has become an increasingly
active subject of research, both from experimental and
theoretical points of view (For recent reviews see \cite{hdqcd}). In
deep inelastic scattering (DIS) the high parton density regime
corresponds to the small $x$ region and represents the challenge of studying the interface between the
perturbative and nonperturbative QCD, with the peculiar feature that this transition is
taken in a kinematical region where the strong coupling constant $\alpha _{s}$ is
small. By the domain of perturbative QCD we mean the region where the parton picture
has been developed and the separation between the short and long distance contributions
(the collinear factorization) is made possible by the use of the operator product
expansion (OPE).  The Dokshitzer-Gribov-Lipatov-Altarelli-Parisi (DGLAP) equations
\cite{dglap},  which resums terms of the type $\alpha_s^n \ln^n Q^2$, are the evolution equations in this kinematical region. These equations
are valid at leading twist,  i.e. at a large value of the photon virtuality $Q^2$, where
a  subclass of all possible Feynmann graphs are dominant and the hard coefficient function 
is connected to the proton by only two parton
lines (For more details see e.g. Ref. \cite{cteq}). For small values of $Q^2$, this picture has corrections predicted by  the OPE 
that 
contribute at relative order ${\cal{O}}(1/Q^2)$ and beyond
$[{\cal{O}}(\frac{1}{Q^2})^n, \,n=2,3,...]$. These are commonly called higher twist
corrections.

In the limit of small values of $x$, terms of the type $\alpha_s^n \ln^n 1/x$ become relevant and  the DGLAP equation is not expected to be valid. In this kinematical region  one expects to see
new features inside the proton: the density of gluons and quarks becomes very high and
an associated new dynamical effect  is expected to stop the
further growth of the structure functions. In particular, for a fixed hard scale $Q^2
\gg \Lambda_{QCD}^2$, the OPE eventually breaks down at sufficiently small $x$
\cite{mueplb}. Ultimately, the physics in the region of high parton densities will be
described by nonperturbative methods, but this is still waiting for a satisfactory
solution in QCD. However, the transition from the moderate $x$ region towards the small
$x$ limit may possibly be accessible in perturbation theory, and, hence, allows us
to test the ideas about the onset of nonperturbative dynamics. Currently, this transition is quite well described by the Color Glass Condensate (CGC) formalism, whose central result is the Jalilian-Marian-Iancu-McLerran-Weigert-Leonidov-Kovner (JIMWLK) equation \cite{CGC,CGC1,CGC2,CGC3,CGC4,CGC5,CGC6,CGC7}. When applied to the scattering
between a simple projectile and a CGC, one obtains an infinite hierarchy of
coupled equations for the correlators of  Wilson lines, the Balitsky-JIMWLK (B-JIMWLK) hierarchy \cite{BAL,BAL1,BAL2,BAL3,BAL4}. In the mean field
approximation, this hierarchy reduces to a single closed nonlinear equation, the Balitsky-Kovchegov (BK) equation \cite{BAL,KOVCHEGOV,KOVCHEGOV1}, which
describes the evolution with energy of the dipole-hadron scattering amplitude
in the large $N_c$ limit ($N_c$ is the number of colors). The CGC physics is characterized by a saturation scale $Q_s$ which is related to the critical transverse size for the unitarization of the scattering amplitudes and is predicted by the BK equation to be an increasing function of the energy and atomic number dependent [$Q_s^2 =  Q_0^2 \, (\frac{x_0}{x})^{\lambda}\,A^{\alpha}$ with $\lambda$ and $\alpha > 0$].  Currently, the  analysis of HERA and RHIC data using phenomenological models based on CGC physics implies that $Q_s \approx 1 - 3$ GeV for proton and nuclei  \cite{hdqcd}. As this scale is not very large at these machines, contributions from nonperturbative dynamics cannot be completely disregarded before estimating its magnitude.



The analytical treatment of the nonperturbative QCD dynamics is one of main
open questions of the strong interactions. Currently, it is only possible
 to get some qualitative information about this regime considering other
 QCD-like theories, which are accessible at strong coupling and share some
 of the basic features of QCD at high energy. One such theory is  the $N=4$
 super Yang-Mills (SYM) where one can perform calculations in the
 nonperturbative  limit of large 't Hooft coupling using the Anti-de Sitter
 space/Conformal field theory (AdS/CFT) correspondence \cite{maldacena,polyakov,witten}
  (For a recent review see \cite{iancu_review}). { However, it should be emphasized 
that QCD is a confining theory with a dynamically generated scale $\Lambda \approx 200$
 MeV, whereas $N=4$ SYM is a conformal theory with no scales and thus does not have confinement.
 Moreover, in $N=4$ SYM there is no mass gap in the asymptotic particle spectrum,
 which implies that the Froissart bound is not expected to be valid. A correspondence
 closer to QCD can be obtained if the conformal invariance of the theory is broken.
 The simplest way to break conformal invariance is by introducing a hard cut-off in 
AdS space  which can be interpreted as an infrared scale of the gauge theory. Another
 possibility is to break conformal invariance softly through a background dilaton 
field. These approaches have been used in last years by several authors in order to 
study the DIS and calculate the high energy scattering amplitudes in the AdS/CFT 
formalism  \cite{polchinski,cornalba,braga,iancu_ads1,iancu_ads2,levin,pire,iancu_ads3,hassa}. { In general, these authors treat the DIS off a shockwave considering the interaction of a {\it virtual photon} with the target.
 In contrast, in \cite{kov_nuc}, 
DIS with a nucleus target was considered in the framework of the shockwave
 approximation with the Wilson loop immersed into a gauge field background  calculated
 following the proposal by Janik and Peschanski \cite{janik}. These authors have considered that the projectile is  a {\it color dipole}, as represented by a Nambu-Goto string.}  This model  was confronted with the available HERA data in Ref. \cite{amir}.
 The basic idea was to construct a dipole-target cross section inspired in the AdS/CFT
 approach valid at small values of the photon virtuality $Q^2$, which can be viewed as
 complementary to the perturbative descriptions based on CGC physics.} A surprising
 prediction of this AdS/CFT inspired model is the $x$ independence of the $F_2$ structure
 function at very small $x$  and $Q^2$  in a region where there is no experimental data
 yet, which is directly associated to the behavior of the saturation scale at high energies.
 This result can be contrasted with the prediction of the high density QCD approaches, where
 $F_2 \propto \ln (1/x)$ in the asymptotic regime \cite{vic_asymp}. Although the relevance
 of the AdS/CFT results for our real QCD  world is far from being clear, the fact that this
 model is able to describe the $F_2$  HERA data in the kinematical region of
 $ x < 6 \times 10^{-5}$ and $Q^2 < 2.5$ GeV$^2$ \cite{amir} motivates one to check what
 are the predictions of this model for other observables. In particular, it is very
 important to determine the signatures of the AdS/CFT inspired model  in order to discriminate
 unambiguously from other approaches, as for example the CGC formalism.

Recently, the running coupling corrections to BK equation were calculated
 through the ressumation of $\alpha_s N_f$
contributions ($N_f$ being the number of flavors) to all orders, allowing the estimation of the soft gluon emission and running coupling corrections to the evolution kernel \cite{kovw1,javier_kov,balnlo,kovw2}. The solution of this improved BK equation was studied in detail in Refs. \cite{javier_kov,javier_prl}.
More recently, a global analysis of the small $x$ ($x \le 10^{-2}$) data  for the proton structure function in the kinematical range $0.045$ GeV$^2$ $\le Q^2 \le 800$ GeV$^2$ 
using the improved BK equation was performed \cite{bkrunning} (See also Ref. \cite{weigert}). 
In contrast to the  BK  equation at leading order (LO), which  fails to describe data, the inclusion
of running coupling effects to evolution renders BK equation compatible with them. In particular, the $F_2$ data in the region of very small $Q^2$ is also quite well described by the improved BK equation. It implies that the current $F_2$ data at low $Q^2$ and $x$ is not able to discriminate between the {\it nonperturbative} AdS/CFT inspired model and the {\it perturbative} CGC physics.

In this paper we analyse the predictions of the AdS/CFT inspired model for  diffractive processes and compare its results with those from the running coupling Balitsky-Kovchegov (RC BK) solution obtained in Ref. \cite{bkrunning}. Our main motivation comes from the fact that in diffractive deep inelastic scattering (DDIS), mainly on $F_2^D$, the interplay between hard and soft regimes is more explicit \cite{Mueller,DDIS}. Basically, the partonic fluctuations of the virtual photon can lead to configurations of
different sizes when analysed in the proton rest frame. The study
of the diffractive dissociation of protons has shown that for real
photons ($Q^2 \approx 0$), where the transverse size of the
incoming pair is approximately that of a hadron, the energy
dependence is compatible with the expectations based on the  soft
Pomeron exchange. On the other hand, at large $Q^2$ the energy
dependence is higher than that of the soft Pomeron, suggesting
that pQCD effects may become visible for small incoming
quark-antiquark pairs. Therefore, as the $F_2^D$ structure
function is inclusive to the hard and soft contributions to the
dynamics, its study can be useful to discriminate between the RC BK and  AdS/CFT predictions.

This paper is organized as follows. In Section II we present the dipole formalism, which describes  the inclusive and diffractive deep
inelastic scattering in a common approach. Within this formalism, we
present the parameterization for the scattering amplitude inspired in the
AdS/CFT correspondence proposed in \cite{kov_nuc}, which has been recently
used to describe the HERA data for the $F_2$ proton structure function
\cite{amir}. Moreover, we analyze the $\rr$ and $x$ - dependence of the scattering amplitude predicted by the  AdS/CFT inpired model and compare it with the RC BK solution. In Section III we analyze the predictions of these models for the inclusive and diffractive overlap functions and calculate the diffractive structure function for small $Q^2$. For completeness, the $F_2$ structure function is also calculated for  similar values of $Q^2$. 
Finally, we calculate the ratio between the diffractive and inclusive total cross sections, $R_{\sigma}=\sigma^D_{tot}/\sigma_{tot}$, and study its energy dependence considering the AdS/CFT inspired model and the RC BK solution.  The conclusions are presented in Section IV.

\section{Dipole formalism}

In the study of the observables of lepton-hadron deep inelastic scattering (DIS) at small $x$, it is convenient to see the scattering between the virtual photon (which is exchanged in the lepton-hadron interaction) and the hadron in the dipole frame:  in this frame, most of the total energy is carried by the hadron, but the virtual photon has enough energy to dissociate into a quark-antiquark ($q\bar q$) pair, a dipole, before the scattering.
In such special frame, the QCD description of DIS at small $x$ can be
interpreted as a two-step process \cite{dipole}: the virtual photon $\gamma^*$ splits into the $q \bar{q}$ dipole, with transverse separation $\rr$, which subsequently interacts with the hadron $h$. In terms of cross sections for the transversely ($T$) and longitudinally ($L$) polarized photons, the $F_2$ structure function is given by $
		F_2(x,Q^2)\,=\,\frac{Q^2}{4 \pi^2 \alpha_{em}} \,\sigma_{tot}$, where $\alphaem$ is the electromagnetic coupling constant and $\sigma_{tot}$ is the total (inclusive) $\gamma^*h$ cross section expressed by
\be
\label{eq:sigtot}
\sigma_{tot} = \sum_{i=T,L}\int d^2{\rr}\, dz\, |\Psi_i(\rr,z,Q^2)|^2\,\, \sigma_{dip}(x,\rr).
\ee
The functions $\Psi_{T,L}$ are the light-cone  wave functions of the virtual photon, $z$ is the photon  momentum fraction carried by the quark (for details see  e.g. Ref. \cite{PREDAZZI}) and $\sigma_{dip}$ is the dipole-hadron cross section.

Diffractive
processes in deep-inelastic scattering (DDIS) are characterized by the presence of large rapidity gaps in the hadronic final state and are associated to a Pomeron exchange. Within the
framework of the perturbative  QCD (pQCD), the Pomeron is associated with the
resummation of leading logarithms in $s$ (center of mass  energy squared) and
at  lowest order is described by the two-gluon exchange. These processes are 
of particular interest,
because the hard photon in the initial state gives rise to  the hope that, at
least in part, the scattering amplitude can be calculated in pQCD.   In the dipole approach  the total diffractive cross sections take the following form  (See e.g. Refs. \cite{GBW,PREDAZZI,dipole}),
\begin{equation}\label{eq:sigdiff}
\sigma^D_{T,L} = \int_{-\infty}^0 dt\,e^{B_D t} \left. \frac{d \sigma ^D _{T,L}}{d t} \right|_{t = 0} = \frac{1}{B_D} \left. \frac{d \sigma ^D _{T,L}}{d t} \right|_{t = 0}
\end{equation}
where 
\begin{equation}\label{eq:dsig-dt}
\left. \frac{d \sigma ^D _{T,L}}{d t} \right|_{t = 0} = \frac{1}{16 \pi} \int d^2 {\bf r} 
\int ^1 _0 d \alpha |\Psi _{T,L} (\alpha, {\bf r})|^2 \sigma _{dip} ^2 (x, \rr) \,\,,
\end{equation}
and it is assumed a factorizable dependence on momentum transfer $t$, on which the
dependence is given through an exponential with diffractive slope $B_D$. 
The diffractive processes can be analyzed in more detail by studying the behavior of the diffractive structure function $F_2^{D (3)}(Q^{2}, \beta, x_{I\!\!P})$. Following Ref. \cite{GBW2} we assume that the diffractive structure function is given by
\begin{equation}
F_2^{D (3)} (Q^{2}, \beta, x_{I\!\!P}) = F^{D}_{q\bar{q},L} + F^{D}_{q\bar{q},T} + F^{D}_{q\bar{q}g,T},
\label{soma}
\end{equation}
where the $q\bar q g$ contribution with longitudinal polarization is not
present because it has no leading logarithm in $Q^2$. The different contributions can be calculated and for the $q\bar q$ contributions
they read \cite{wusthoff,GBW2,dipole}
\begin{equation}
  x_{I\!\!P}F^{D}_{q\bar{q},L}(Q^{2}, \beta, x_{I\!\!P})=
\frac{3 Q^{6}}{32 \pi^{4} \beta B_D} \sum_{f} e_{f}^{2} 
 2\int_{\alpha_{0}}^{1/2} d\alpha \alpha^{3}(1-\alpha)^{3} \Phi_{0},
\label{qqbl}
\end{equation}
\begin{equation}
 x_{I\!\!P}F^{D}_{q\bar{q},T}(Q^{2}, \beta, x_{I\!\!P}) =  
 \frac{3 Q^{4}}{128\pi^{4} \beta B_D}  \sum_{f} e_{f}^{2} 
 2\int_{\alpha_{0}}^{1/2} d\alpha \alpha(1-\alpha) 
\left\{ \epsilon^{2}[\alpha^{2} + (1-\alpha)^{2}] \Phi_{1} + m_f^{2} \Phi_{0}  \right\}   
\label{qqbt}
\end{equation}
where the lower limit of the integral over $\alpha$ is given by $\alpha_{0} = \frac{1}{2} \, \left(1 - \sqrt{1 - \frac{4m_{f}^{2}}{M_X^{2}}}\right)
$, the sum is performed over the quark flavors and \cite{fss}
\begin{equation}
\Phi_{0,1}  \equiv  \left(\int_{0}^{\infty}r dr K_{0 ,1}(\epsilon r)\sigma_{dip}(x_{I\!\!P},\rr) J_{0 ,1}(kr) \right)^2.
\label{fi}
\end{equation}
The $q\bar{q}g$ contribution, within the dipole picture at leading $\ln Q^2$ accuracy, is given by \cite{wusthoff,GBW2,nikqqg,golec}
 \begin{eqnarray}
   \lefteqn{x_{I\!\!P}F^{D}_{q\bar{q}g,T}(Q^{2}, \beta, x_{I\!\!P}) 
  =  \frac{81 \beta \alpha_{S} }{512 \pi^{5} B_D} \sum_{f} e_{f}^{2} 
 \int_{\beta}^{1}\frac{\mbox{d}z}{(1 - z)^{3}} 
 \left[ \left(1- \frac{\beta}{z}\right)^{2} +  \left(\frac{\beta}{z}\right)^{2} \right] } \label{qqg} \\
  & \times & \int_{0}^{(1-z)Q^{2}}\mbox{d} k_{t}^{2} \ln \left(\frac{(1-z)Q^{2}}{k_{t}^{2}}\right) 
\left[ \int_{0}^{\infty} u \mbox{d}u \; \sigma_{dip}(u / k_{t}, x_{I\!\!P}) 
   K_{2}\left( \sqrt{\frac{z}{1-z} u^{2}}\right)  J_{2}(u) \right]^{2}.\nonumber
\end{eqnarray} 
As pointed in Ref. \cite{dipolos9}, at small $\beta$ and low $Q^2$, the leading $\ln (1/\beta)$ terms should be resumed and the above expression should be modified. However, as a description with the same quality using the Eq. (\ref{qqg}) is possible by adjusting the coupling \cite{dipolos9}, in what follows we will use this expression for our phenomenological studies. 
We  use the standard notation for the variables $\beta = Q^2 / (M_X^2 + Q^2)$, $
x_{I\!\!P} = (M_X^2 + Q^2)/(W^2 + Q^2)$ and $x = Q^2/(W^2 + Q^2) = \beta x_{\pom}$, 
where $M_X$ is the invariant mass of the diffractive system, $B_D$ is the diffractive slope and $W$ the total energy of the 
$\gamma ^* p$ system.


\begin{figure}[t]
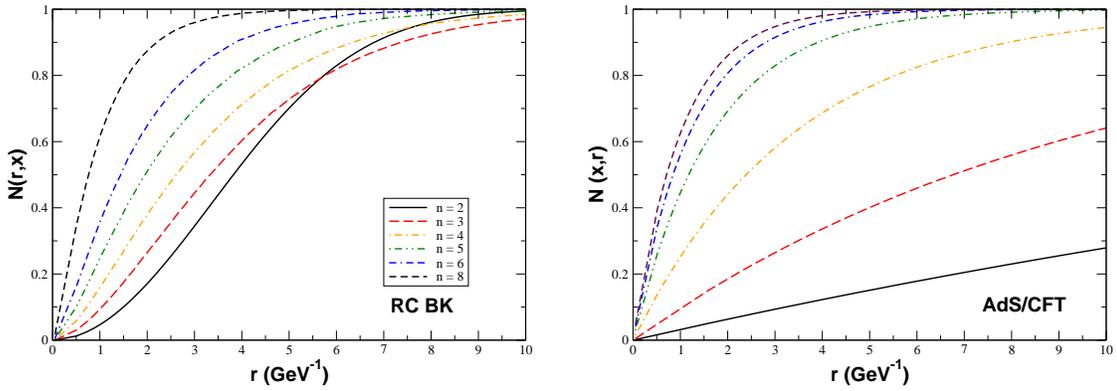

\centerline{
{\psfig{figure=ene_bknlo.eps,width=7.cm}}
\hspace{0.5cm}
{\psfig{figure=enes_ads.eps,width=7.cm}}}
\caption{Dependence of the dipole scattering amplitude in the pair separation $r$ at 
different values of $x$ ($x = 10^{-n}$).}
\label{fig1}
\end{figure}

The main input for the calculations of  inclusive and diffractive observables in the dipole picture is $\sigma_{dip}(x,\rr)$ which is determined by the QCD dynamics at small $x$. In the eikonal approximation, it is  given by:
\begin{equation} 
\sigma_{dip}(x, \rr) = 2 \int d^2 \rb \,  {\cal N}(x, \rr, \rb)
\label{sdip}
\end{equation}
where $ {\cal N}(x, \rr, \rb)$ is the forward scattering amplitude for a dipole with size 
$r=|\rr|$ and impact parameter $\rb$  which can be related to expectation value of a Wilson loop \cite{hdqcd}. It
encodes all the
information about the hadronic scattering, and thus about the
non-linear and quantum effects in the hadron wave function. In general, it is  assumed that the impact parameter dependence of $\cal{N}$ can be factorized as  ${\cal{N}}(x,\rr,\rb) = {\cal{N}}(x,\rr) S(\rb)$, where
$S(\rb)$ is the profile function in impact parameter space, which implies  $\sigma_{dip}(x,\rr)=\sigma_0 \mathcal{N}(x,\rr)$. The forward scattering amplitude ${\cal{N}}(x,\rr)$ 
can be obtained by solving the BK evolution equation \cite{bkrunning} or considering phenomenological QCD inspired models to describe the interaction of the dipole with the target \cite{amir,GBW,GBW2,dipolos2,dipolos3,dipolos4,dipolos5,dipolos6,dipolos7,dipolos8,
dipolos9,dipolos10,iim,kkt,dhj,dhj1,Goncalves:2006yt,buw,fks}. BK equation is the simplest nonlinear evolution equation
for the dipole-hadron scattering amplitude, being actually a mean field version
of the first equation of the B-JIMWLK hierarchy \cite{BAL,BAL1}. In its linear
version, it corresponds to the Balitsky-Fadin-Kuraev-Lipatov (BFKL) equation
\cite{bfkl}.
The LO BK equation presents some difficulties when applied to study
DIS small-$x$ data, in particular, some studies concerning this equation
\cite{IANCUGEO,MT02,AB01,BRAUN03,AAMS05} have
shown that the resulting saturation scale grows much faster with increasing energy
($Q_s^2\sim x^{-\lambda}$, with $\lambda \approx 0.5$) than that
extracted from phenomenology ($\lambda \sim 0.2-0.3$). This difficulty could be solved by
considering higher order corrections to LO BK equation, which were recently calculated \cite{kovw1,javier_kov,balnlo,kovw2} and have shown to successfully
describe small-$x$ ($x \le 10^{-2}$) data  for the proton structure function 
\cite{bkrunning}. In what follows we will use in our RC BK calculations the public-use code available in \cite{code}.




\vspace{0.5cm}
\begin{figure}[t]
\centerline{
{\psfig{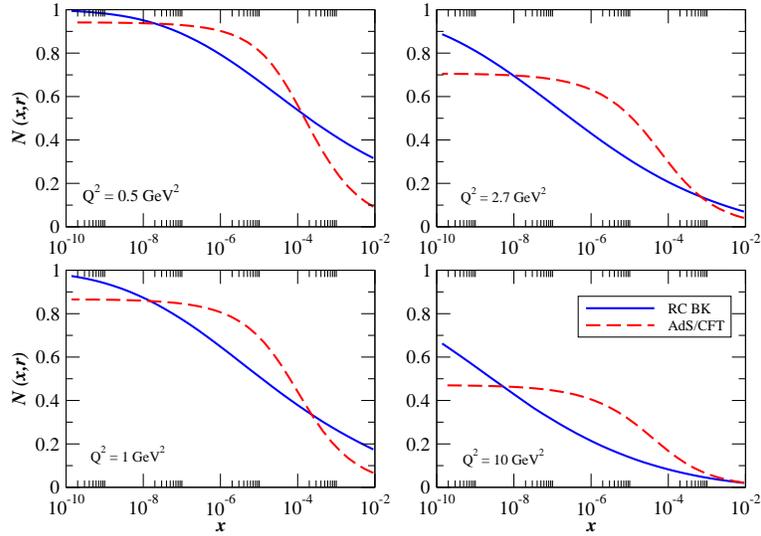}}}
\caption{Energy dependence of the scattering amplitude as a function of $x$ at different values of $Q^2$ ($r^2 = 1/Q^2$).}
\label{fig2}
\end{figure}


An alternative approach is to calculate the scattering amplitude at high energy
for a strongly coupled $N = 4$ super Yang-Mills theory using the AdS/CFT
correspondence. In  \cite{kov_nuc} the authors have modelled the nucleus
by a metric of a shock wave in AdS$_5$ and calculated the total cross section 
or DIS on a large nucleus. The expectation value of the Wilson loop, which is
directly related to $\cal{N}$, was calculated by finding the extrema of the
Nambu-Goto action for an open string attached to the quark and antiquark lines
of the loop   in the background of an AdS$_5$ shock wave. The resulting scattering
amplitude  can be expressed in terms of a AdS/CFT saturation scale, which is energy
independent at very small $x$ and depends very strongly on the atomic number of
the nucleus ($\propto A^{1/3}$). In
\cite{amir} it was  used to describe the HERA data for $F_2$ proton structure function
in the small $(x,Q^2)$ range  and a parametrization for the dipole-proton scattering
amplitude was provided.  In this AdS/CFT inspired model the scattering amplitude is
given by \cite{kov_nuc}
\begin{equation}\label{eq:Nads}
{\cal{N}}(r,x)=1-\exp\left[-\frac{\mathcal{A}_0xr}{\mathcal{M}_0^2(1-x)
\pi\sqrt{2}}\left(\frac{1}{\rho_m^3}+
\frac{2}{\rho_m}-2\mathcal{M}_0\sqrt{\frac{1-x}{x}}\right)\right]
\end{equation}
with 
\begin{eqnarray}
\rho_m&=&  \begin{cases}
 (\frac{1}{3m})^{1/4}\sqrt{2\cos(\frac{\theta}{3})}  & : m \le \frac{4}{27} \nonumber\\
     \sqrt{\frac{1}{3m\Delta} +  \Delta} & :m> \frac{4}{27}
  \end{cases},  \nonumber\\
\Delta&=& \Big[\frac{1}{2m}-\sqrt{\frac{1}{4m^2}-\frac{1}{27m^3}}\Big]^{1/3} \nonumber\\
m&=&\frac{\mathcal{M}_0^4(1-x)^2}{x^2}, \nonumber\\
\cos(\theta)&=&\sqrt{\frac{27m}{4}}. \label{not}\
\end{eqnarray}
In the above equations, $\mathcal{M}_0=b_0c_0$, where $b_0$ relates the
virtuality of the photon to the dipole size, $b_0=rQ$, and
$c_0=\Gamma^2(\frac 1 4)/(2\pi)^{3/2} \approx 0.83$; this is the value
predicted by AdS/CFT calculations, however, from the fit to HERA data it
assumes a value about 2 orders of magnitude smaller, which may be explained
by the fact that $N=4$ SYM theory and QCD are diferent from each
other (see the discussion in Section IV of Ref.\cite{amir}).
$\mathcal{A}_0=\sqrt{\lambda_{\rm{YM}}}\Lambda$, where
$\lambda_{\rm{YM}}=g^2_{\rm{YM}}N_c$ denotes the 't Hooft coupling with
$g_{\rm{YM}}$ the Yang-Mills coupling constant, and $\Lambda$ is chosen to
be $\Lambda=1$ GeV. The resulting saturation scale  has the
following form:
\begin{equation} 
Q_s^{\text{AdS}}(x)=\frac{2 \, \mathcal{A}_0 \, x}{\mathcal{M}_0^2 \, (1-x) \, \pi} \, 
\left(\frac{1}{\rho_m^3}+\frac{2}{\rho_m}-2\mathcal{M}_0\sqrt{\frac{1-x}{x}}\right).
\label{qads}
\end{equation}
It has its values
in the interval 1 \textdiv 3 GeV in the range $6.2\times 10^{-7}\leq x \leq 6\times 10^{-5}$ and has a
particular property: for extremely small values of $x$, the saturation scale saturates, which is in sharp contrast with its usual form in QCD, where
it grows with decreasing $x$. This behavior implies that the DIS cross sections will present a slow growth with the energy, similar to that predicted
by soft pomeron models \cite{kov_nuc}.
In what follows we use $m_f=140$ MeV for the (light) quark masses and the other
parameters are chosen based on the analysis done in \cite{amir}; in particular,
the value of the t'Hooft coupling constant is chosen to be
$\lambda_{\rm{YM}}=20$: this choice is consistent with the fact that AdS/CFT
correspondence is valid for $\lambda_{\rm{YM}}\gg 1$ and motivated mainly
because, besides the resulting good $\chi^2$, the results of fit to HERA data
change only a little bit for a wide range $\lambda_{\rm{YM}}\geq 20$. This gives
$\mathcal{M}_0=6.54\times 10^{-3}$ and $\sigma_0=22.47$ mb.


In Fig. \ref{fig1} we show the pair separation dependence of the scattering amplitude $\cal{N}$ for different values of $x$ considering the AdS/CFT inspired model and the RC BK solution (For a related discussion see Refs. \cite{testing,nos}).  Although the  AdS/CFT model is constructed in order to describe small $x$ ($\le 10^{-4}$) we also present its predictions at larger $x$ in order to compare with the RC BK solution. The first aspect that can be observed is the large difference between the predictions at  $x > 10^{-4}$. While the RC BK solution predicts that the scattering amplitude saturates at large $x$, the AdS/CFT  model still predicts a growth  in the $r$-range considered. In the range  $ 10^{-6} \le x \le  10^{-4}$ the behavior predicted by the two models for $\cal{N}$ is similar, which is expected since it is the range of the HERA data used to constrain the main parameters of the AdS/CFT model. In this region both models predict that $\cal{N}$ saturates at large pair separation. At smaller values of $x$, the two models predict different behaviors.  While the scattering amplitude in the AdS/CFT model is almost $x$ independent, the RC BK solution is significantly modified when $x$ goes  to smaller values.  In particular, at $x \le 10^{-8}$ the  scattering amplitude in the  AdS/CFT inspired model becomes energy independent. It can be observed in more detail in Fig. \ref{fig2}, where we analyze the $x$ dependence of the scattering amplitude for different values of $Q^2$, i.e. different values of the squared pair separation if we assume $r^2 = 1/Q^2$. As expected from Fig. \ref{fig1}, the two predictions are very distinct at large $x$ ($ \ge 10^{-4}$) and large pair separations ($Q^2 = 0.5$ and 1 GeV$^2$), being smaller for larger values of $Q^2$. In contrast to the RC BK solution, which predicts that
$\cal{N}$ grows at small $x$ and large $Q^2$, the AdS/CFT model predicts that $\cal{N}$ saturates at $x \le 10^{-6}$ independently of the pair separation. As in the dipole approach the  observables are determined by the scattering amplitude, we can expect that these differences between the models also be observed in the inclusive and diffractive structure functions.


\vspace{0.5cm}
\begin{figure}[t]
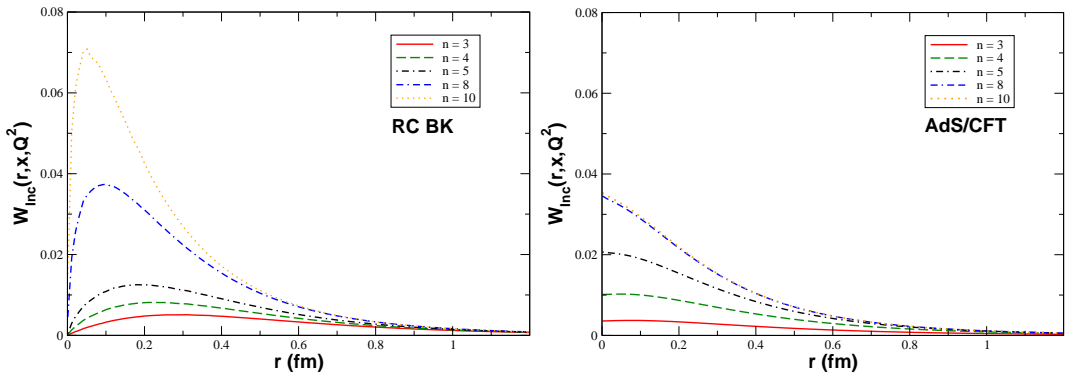

\centerline{
{\psfig{figure=weight_inc_bk.eps,width=7.cm}}
{\psfig{figure=weight_inc_ads.eps,width=7.cm}}
}
\caption{Pair separation dependence of the function $W_{Inc}(r,x,Q^2)$ for distinct values of $x$ ($x=10^{-n}$) and $Q^2 = 1$ GeV$^2$.}
\label{fig3}
\end{figure}

\section{Results}
\label{res}


In  DIS   the partonic fluctuations of the virtual photon  can lead to
configurations of different sizes when analysed in the dipole frame. The
size of the configuration will depend on the relative transverse momentum
$k_T$ of the $q\overline{q}$ pair. The small size configurations are
calculated using perturbative QCD and at small Bjorken scaling variable $x$ the smallness of the cross section (color transparency) is compensated by the large gluon distribution. For large size configurations one
expects to be in the regime of soft interactions. In the inclusive measurement of final states 
one sums over both small-distance and large-distance configurations. The contribution of small and large pair separations for inclusive and diffractive observables can be studied considering the corresponding overlap functions which are given by:
\begin{eqnarray}
W_{Inc} \,(r,x,Q^2)  = { 2\pi \rrn} \sum_{i=T,L} \, \int dz\, |\Psi_i
(z,\,r, Q^2)|^2 \, \sigma_{dip} (x,r)\, ,
\label{eq:overlap-incl}
\end{eqnarray}
and 
\begin{equation}
W_{Diff} \,(r,x,Q^2)  =  2\pi \rrn \,\sum_{i=T,L} \int d \alpha \,  |\Psi_i
(\alpha,\,r, Q^2)|^2 \, \sigma_{dip}^2 (x,r)\,.
\label{eq:overlap-diff}
\end{equation}
It is important to emphasize that the behavior of the overlap functions are strongly dependent on the scattering amplitude. In  particular, as shown in \cite{GBW}, if one considers a model where ${\cal{N}} \propto r^2$ (linear dynamics) the  diffractive overlap function is dominated by large pair separations and consequently the diffractive cross sections are determined by nonperturbative physics. Moreover, it is the energy dependence of the scattering amplitudes which determine the  $x$ dependence of the overlap functions.

In Fig. \ref{fig3} we present our results for the inclusive overlap function considering the AdS/CFT inspired model and the RC BK solution at different values of $x$ ($x=10^{-n}$ with $n =$ 3, 4, 5, 8, 10) and $Q^2 = 1$ GeV$^2$. We have that both models predict a similar behavior for the large $r$ region, which implies that the main contribution for inclusive observables comes from  small pair separations. As $r \rightarrow 0$ the RC BK solution predicts that $W_{Inc}$ goes to zero, which is directly associated to the BFKL behavior present in the BK equation at very  small $r$. 
In contrast, the AdS/CFT inspired model predicts that in this limit $W_{Inc}$ goes to a constant value, which is energy-dependent. Another aspect which should be emphasized is that the inclusive overlap becomes energy independent at $x \le 10^{-8}$ for the AdS/CFT model, in agremment with the behavior of the corresponding scattering amplitude. In contrast,  
RC BK solution implies that $W_{Inc}$ is energy dependent. These behaviors have direct impact in  the predictions for the  $F_2$ structure function, as will be verified below.  Finally, the peak of the  inclusive overlap function occurs at smaller values of $r$ with the decreasing of $x$.

\begin{figure}[t]
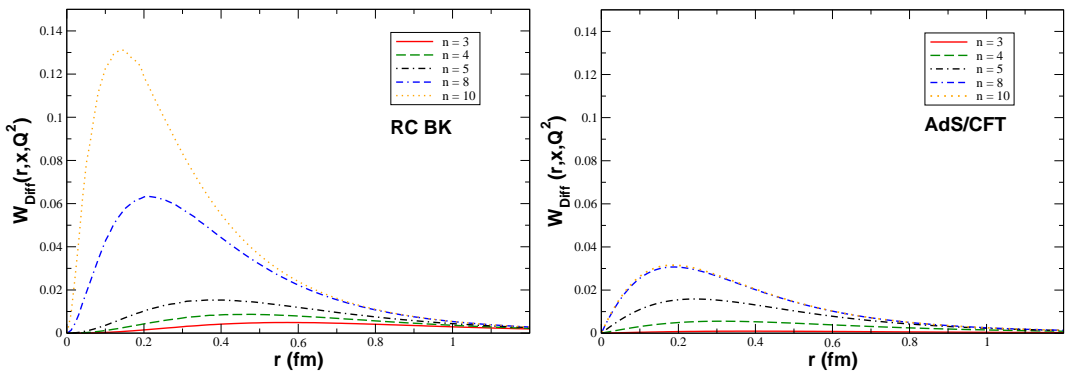

\vspace*{1.5cm}
\centerline{
{\psfig{figure=weight_dif_bk.eps,width=7.cm}}
{\psfig{figure=weight_dif_ads.eps,width=7.cm}}
}
\caption{Pair separation dependence of the function $W_{Diff}(r,x,Q^2)$ for distinct values of $x$ ($x=10^{-n}$) and $Q^2 = 1$ GeV$^2$.}
\label{fig4}
\end{figure}

As discussed in the Introduction, the main theoretical interest in diffraction is centered around the interplay between the soft and hard physics.  The ability to clearly separate the regimes dominated by soft and hard processes is essential in exploring QCD at both quantitative and qualitative levels. Moreover, as the diffractive observables ($\sigma^D_{T,L}$ for example) are proportional to $\sigma_{dip}^2$, these are more sensitive to the large pair separation contributions \cite{GBW2}.  These features of the diffractive processes make DDIS an ideal scenario to check the AdS/CFT inspired dipole model, since this is a nonperturbative inspired model of saturation. 
In Fig. \ref{fig4} we present our results for the diffractive overlap function considering the AdS/CFT inspired model and the RC BK solution at different values of $x$ ($x=10^{-n}$ with $n =$ 3, 4, 5, 8, 10) and $Q^2 = 1$ GeV$^2$.
We can see that both models predict that the large pair separations do not contribute significantly for the diffractive overlap function. Moreover, $W_{Diff}(r,x,Q^2)$ goes to zero at $r \rightarrow 0$ for both models. The peak of the distribution occurs at smaller values  of $r$ with the decreasing $x$. In comparison with the inclusive case, it peaks at larger values of $r$, which implies that the main contribution for  diffractive observables comes from a different range  of values of pair separation and consequently probe distinct aspects of the QCD dynamics present in the scattering amplitude.
 The small $x$ effects in the scattering amplitude predicted by the AdS/CFT  inspired model and the RC BK solution  drive the energy behavior of the diffractive overlap function. In Fig. \ref{fig2} the $x$-dependence of the ${\cal N}(x,r)$ shows that for  $x<10^{-6}$ the AdS/CFT predicts a saturation of the amplitude, while the RC BK predicts a saturation only for $x<10^{-10}$. This is one of the reasons for the distinct normalization of the overlap function. The saturation of the AdS/CFT dipole model for small $x$ is more evident if one compares the overlap function for $x=10^{-8}$ and $x=10^{-10}$, since the curves are identical. The AdS/CFT dipole amplitude presents smaller values when compared to the RC BK model in the region $x\gsim 10^{-4}$ and $x \lsim 10^{-6}$. This explains the different normalizations of the overlap functions in the Fig. \ref{fig4}.



\begin{figure}[t]
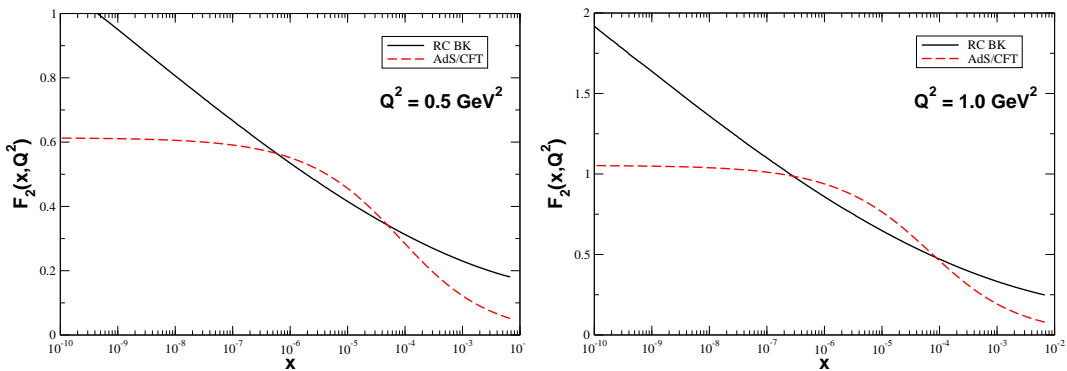

\vspace*{1.5cm}
\centerline{
{\psfig{figure=f2_q2_0_5.eps,width=7.cm}}
{\psfig{figure=f2_q2_1.eps,width=7.cm}}
}
\caption{The $F_2$ structure function for two distinct values of $Q^2$.}
\label{fig5}
\end{figure}

Having addressed  the main features of the overlap functions, we continue
by studying the inclusive and diffractive proton structure functions. We focus our analysis in the small $(x,Q^2) $ kinematical range probed by HERA and which could be tested in future $ep$ colliders at the TeV energy scale, as for example the LHeC project (For a recent discussion see \cite{lhec}). In Fig. \ref{fig5} we show the $F_2$ proton structure function as a function of $x$ for two different small values of the photon virtuality, $Q^2=0.5$ and 1 GeV${}^2$, using the AdS/CFT inspired model (dashed line) and, for the sake of comparison, the RC BK solution (solid line). One can clearly see that both analyses match only in a restrict region, for values of $x$ between
$\sim 10^{-4}$ and $\sim 10^{-7}$; such an interval
corresponds to the HERA kinematical regime investigated in Ref.\cite{amir}. For values of $x$ larger than $10^{-4}$ and,
specially, smaller than $\sim 10^{-7}$ (for which there are no available
experimental data) the behaviors of the curves become distinct. This distinct behavior of $F_2(x,Q^2)$ is related to the $x$ dependence of the scattering amplitude in the AdS/CFT and RC BK models, analysed in the Fig. \ref{fig2}.
In particular, we have that the AdS/CFT inspired model predicts the saturation of the $F_2$ structure function at $\lesssim 10^{-6}$. At $x = 10^{-8}$ the predictions differ by a factor 1.4.

\begin{figure}[t]
\vspace*{1.5cm}
\centerline{
{\psfig{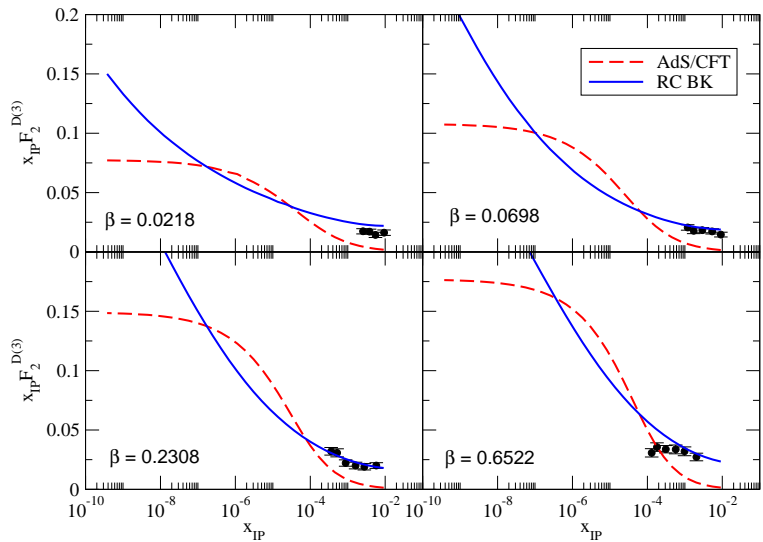}}
}
\caption{Predictions for $F_2^{D (3)} (Q^{2}, \beta, x_{I\!\!P})$  compared with the ZEUS data \cite{zeusdata}. Solid line: RC BK model; Dashed line: AdS/CFT model.}
\label{predictf2d3}
\end{figure}

Let us now discuss the predictions of the AdS/CFT inspired model for the diffractive structure function. As explained before the
main theoretical interest in diffraction is centered around the interplay between the soft and hard physics.
In particular, in the inclusive measurement
of diffractive final states, where the diffractive structure function is
derived, one sums over both small-distance and large-distance configurations.
 As in the diffractive cross
section we integrate over both perturbative and nonperturbative regions of
the phase space, there is a competition between these two pieces. Therefore, in principle, this observable is ideal to test the nonperturbative physics present in the AdS/CFT inspired model.
Before the comparison of the predictions with the HERA data a comment is in order. The diffractive structure function has been measured by the H1 and ZEUS Collaborations in $ep$ collisions at HERA in a large kinematical ($Q^2, x$) range. However, as the model proposed in \cite{amir} is only valid at small $Q^2$ we will select the HERA data in this range. This limits our comparison to the ZEUS data at $Q^2 = 2.7$ GeV$^2$ \cite{zeusdata}, which is similar to the upper limit considered in \cite{amir}. 
In Fig. \ref{predictf2d3} we compare the predictions of the AdS/CFT inspired model with the $F_2^{D(3)}$ ZEUS data for different values of $\beta$. For comparison, we also present the RC BK prediction, which was recently analyzed in detail in  the Ref. \cite{nos} and a good description of the  data in a large $Q^2$ range  was found. The  additional parameters in our calculations using the AdS/CFT inspired model are the
strong coupling constant, which determines the normalization of the $q\bar{q}g$ contribution, and 
 the diffractive slope $B_D$. Following Ref. \cite{nos} we assume  $\alpha_s=0.15$ and constrain the diffractive
slope $B_D$  by the value of $\sigma_0$ through $\sigma_0 =4\pi B_D$ assuming a Gaussian form factor for the proton \cite{dipolos9}. As already demonstrated in \cite{nos} the RC BK solution describes reasonably well the data. In contrast, 
the AdS/CFT inspired predictions underestimate the ZEUS data, which is in the kinematical range $x_{I\!P} \gtrsim 10^{-4}$. For $10^{-7}\lsim x_{I\!P} \lsim 10^{-4}$ is hard to distinguish between the parametrizations. However, for $x_{I\!P} \lsim 10^{-7}$ and all $\beta$ the AdS/CFT inspired model predict that the diffractive structure function saturates, while the RC BK solution predicts a steep growth at small  $x_{I\!P}$. This behavior is directly  related to the energy behavior of the dipole scattering amplitude shown in Figs. \ref{fig1} and \ref{fig2}. At  $x_{I\!P} = 10^{-8}$ the predictions differ by a factor 1.5.

\begin{figure}[t]
\vspace*{1.5cm}
\centerline{
{\psfig{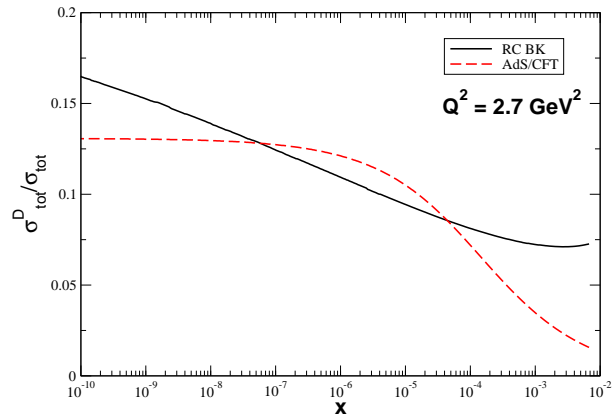}}
}
\caption{Prediction for the $x$  dependence of the ratio $\sigma^D_{tot}/\sigma_{tot}$.}
\label{fig7}
\end{figure}

Finally, in Fig. \ref{fig7} we present the energy dependence of the  ratio $R_{\sigma}=\frac{\sigma^D_{tot}}{\sigma_{tot}}$, which has been considered an important signature of the saturation physics \cite{GBW2,simoneDiff}.
In the range $10^{-7} \lsim\,x\,\lsim 10^{-4}$  both models predict a similar behavior. At $x_{I\!P}\lsim 10^{-7}$ the models predicts distincts behaviours, being $R_{\sigma}$ energy independent for $x\,\lsim\, 10^{-7}$ from the AdS/CFT model. This behavior  is directly associated to the $x$ dependence of the dipole amplitude, predicted by the models (see Fig. \ref{fig2}). Therefore, only at $x\,\lsim\,10^{-7}$ one expects to find some distinction between the parametrizations for this observable, which means that only at very small $x$ we expect some difference between perturbative and strong coupling models.

\section{Conclusions}

Disentangle  the hard and soft dynamics in diffractive DIS is one of the main
open questions of the strong interactions. Although Regge theory is able to
reproduce the experimental data for hadronic collisions, the understanding of
this theory in terms of QCD is still lacking, which is directly associated to
the fact that strong coupling analytical calculations are not possible in QCD.
It is expected that considering the high energy scattering amplitude in $N = 4$
SYM theory using the AdS/CFT correspondence one can guess which physics
phenomena could be important in the strong coupling limit of QCD.
In this paper we have investigated inclusive and diffractive lepton-hadron
DIS, within the dipole formalism, through a parametrization for the
dipole-hadron scattering amplitude presented recently in \cite{kov_nuc},
inspired on the AdS/CFT correspondence, proposed to describe the evolution
and scattering at small values of $x$ and $Q^2$,
i.e., proposed to describe a nonperturbative domain, characterized by large
values of the coupling constant.  We have compared the results with a
perturbative QCD-based framework where the scattering amplitude is the solution
of the nonlinear BK evolution equation with running coupling effects
\cite{bkrunning}. Both analyses are similar only in a restrict kinematical
range ($10^{-7}\lsim \,x \,\lsim 10^{-4}$), and become quite different
when $x$ goes towards large (where the AdS/CFT dipole model is not valid)
and extremely small values, which are not experimentaly available yet.
In particular, the diffractive structure function $F_2^{D(3)}$, which
has been shown to be well described by the solution of the running
coupling BK equation \cite{nos}, is underestimated by the analysis
using the AdS/CFT dipole model in the restricted range of available experimental
data. { Our results indicate that the strict AdS/CFT approach behind
the dipole model proposed in \cite{kov_nuc}, could not be the right
approach for the treatment of diffractive deep inelastic
scattering in the HERA kinematical range.} Moreover, discriminating between the
perturbative RC BK approach and the nonperturbative AdS/CFT inspired approach
in the kinematical range of the future $ep$ colliders \cite{lhec} will be a hard task.

The noticeable difference between RC BK and AdS/CFT analyses at extremely small
values of $x$ is due to the asymptotic prediction of the latter, i.e.,
the saturation scale \textit{saturates} in the $x\rightarrow 0$ limit,
and it is a characteristic of the particular  AdS/CFT model derived in \cite{kov_nuc}. 
{It must be pointed out, that, besides surprising, this behavior of $Q_s$ is also controversial,
as it was argued in \cite{iancu_ads3}. In particular, it is in sharp contradiction with
all other calculations of the scattering amplitudes in the context of the AdS/CFT correspondence,
whose predictions give $Q_s^2 \sim 1/x$, i.e., the saturation scale grows much faster than the
corresponding scale in perturbative QCD.   An energy-dependent saturation scale appears to be consistent with the high
energy dynamics at strong coupling and, as it was demonstrated in \cite{iancu_ads3}, it is
necessary to ensure energy-momentum conservation. As discussed in the Introduction, the basic difference between the  model proposed in \cite{kov_nuc} and the other ones in the literature is the nature of the projectile, which is assumed to be a color dipole in \cite{kov_nuc}, instead of a virtual photon as in  \cite{iancu_ads3}. This difference should be unimportant at a fundamental level, since the property of unitarization refers to the mechanism of inelastic scattering. It implies that, probably, the different energy dependence of the saturation scale should be associated to some subtle mathematical manipulation which were not fully under control.
Finally, if the behavior $Q_s^2 \sim 1/x$ is
the correct one, it can be expected that the resulting AdS/CFT predictions will be closer to
the perturbative BK predictions in the kinematical range of the future colliders. Therefore,
in order to discriminate between the different formalisms, it will be necessary to search other
less inclusive observables, as for example, exclusive vector meson photoproduction.}

\begin{acknowledgments}
This work was  partially financed by the Brazilian funding agencies CNPq, CAPES and
FAPERGS.
\end{acknowledgments}

\bibliographystyle{unsrt}
\bibliography{refs_AdS_CFT2}

\end{document}